# Possible competing order-induced Fermi arcs in cuprate superconductors

B.-L. Yu[*], J. C. F. Wang[*], A. D. Beyer, M. L. Teague, G. P. Lockhart, J. S. A. Horng[*], S.-P. Lee[*], N.-C. Yeh[†]

*Department of Physics, California Institute of Technology, Pasadena, CA 91125, USA*

**Abstract**

We investigate the scenario of competing order (CO) induced Fermi arcs and pseudogap in cuprate superconductors. For hole-type cuprates, both phenomena as a function of temperature and doping level can be accounted for if the CO vanishes at $T^*$ above the superconducting transition $T_c$ and the CO wave-vector **Q** is parallel to the *antinodal* direction. In contrast, the absence of these phenomena and the non-monotonic *d*-wave gap in electron-type cuprates may be attributed to $T^* < T_c$ and a CO wave-vector **Q** parallel to the *nodal* direction.

One of the most debated issues in high-temperature superconductivity is the physical origin of various unconventional and often non-universal phenomena observed at temperatures above the superconducting transition $T_c$ [1-5]. Most of these strongly doping dependent phenomena are only associated with the hole-type cuprates, and are particularly pronounced in the underdoped regime. The specific unconventional phenomena include: opening of a pseudogap (PG) at a temperature $T^* > T_c$, below which there is incomplete suppression of the electronic density of states; formation of the "Fermi arcs" [3,6-8] at $T_c < T < T^*$, which refers to the occurrence of a truncated Fermi surface in the PG state that is intermediate between the node of the $d_{x^2-y^2}$-wave superconducting state at $T < T_c$ and the full Fermi surface of the normal state at $T > T^*$; marginal Fermi liquid behavior that leads to unconventional temperature dependence in the resistivity and magnetic susceptibility [9]; and anomalous Nernst effect above the superconducting transition [10]. Various theoretical models have been proposed to account for these unconventional quasiparticle excitations in the PG state. One school of thought may be generally referred to as the ``one-gap'' or ``preformed pair'' model [1,11-14], which asserts that the onset of pair formation occurs at $T^*$ and that the PG state at $T_c < T < T^*$ is a disordered pairing state with strong phase fluctuations. The other school of thought considers the possibility of competing orders (CO's) [1,2,9,15-19] so that one other energy scale $V_{CO}$ besides the superconducting gap $\Delta_{SC}$ is responsible for the low-energy quasiparticle excitations. To date a number of experimental findings seem to favor this "two-gap" concept [7,20-23], although quantitative analyses of the data based on this scenario were lacking. On the other hand, the preformed-pair model and the CO scenario need not be mutually exclusive. For instance, in the event of CO's coexisting with coherent Cooper pairs in the ground state, there is no reason why they cannot coexist with incoherent preformed pairs in the pseudogap phase slightly above $T_c$.

Recently, we have employed a phenomenological approach to quantitatively investigate how coexisting CO's and superconductivity (SC) may influence the low-energy quasiparticle excitations of the cuprates with different doping levels and for $0 \le T <\sim T_c$ [24-26]. We find that the phenomenology not only accounts for the presence (absence) of the low-energy PG in hole-type (electron-type) cuprate superconductors but also reconciles a number of seemingly non-universal experimental findings [24-26]. The primary objective of this work is to extend our previous studies for $0 \le T <\sim T_c$ to the PG state at $T_c < T < T^*$ in order to explore whether the CO scenario can explain the presence (absence) of the Fermi arcs in hole-type (electron-type) cuprates. We demonstrate below that our CO-scenario can account for experimental results from the angle-resolved photoemission spectroscopy (ARPES) and the scanning tunneling microscopy (STM).

Our approach begins with a mean-field Hamiltonian $\mathcal{H}_{MF} = \mathcal{H}_{SC} + \mathcal{H}_{CO}$ that consists of coexisting SC and a CO at $T = 0$ [24,25]. We further assume that the SC gap $\Delta_{SC}$ vanishes at $T_c$ and the CO order parameter vanishes at $T^*$, and that both $T_c$ and $T^*$ are second-order phase transitions. The SC Hamiltonian is given by:

$$\mathcal{H}_{SC} = \sum_{\mathbf{k},\alpha} \xi_\mathbf{k} c^\dagger_{\mathbf{k},\alpha} c_{\mathbf{k},\alpha} - \sum_\mathbf{k} \Delta_{SC}(\mathbf{k}) \left( c^\dagger_{\mathbf{k},\uparrow} c^\dagger_{-\mathbf{k},\downarrow} + c_{-\mathbf{k},\downarrow} c_{\mathbf{k},\uparrow} \right), \quad (1)$$

---

where $\Delta_{SC}(\mathbf{k}) = \Delta_{SC}(\cos k_x - \cos k_y)/2$ for $d_{x^2-y^2}$-wave pairing, $\mathbf{k}$ denotes the quasiparticle momentum, $\xi_\mathbf{k}$ is the normal-state eigen-energy relative to the Fermi energy, $c^\dagger$ and $c$ are the creation and annihilation operators, and $\alpha = \uparrow, \downarrow$ refers to the spin states. The CO Hamiltonian is specified by the energy $V_{CO}$, a wave-vector $\mathbf{Q}$, and a momentum distribution $\delta\mathbf{Q}$ that depends on a form factor, the correlation length of the CO, and also on the degree of disorder. We have previously considered the effect of various types of CO's on the quasiparticle spectral density function $A(\mathbf{k},\omega)$ and the density of states $\mathcal{N}(\omega)$. Specifically, for charge density waves (CDW) being the relevant CO [2,17], we have a wave-vector $\mathbf{Q}_1$ parallel to the CuO$_2$ bonding direction $(\pi,0)$ or $(0,\pi)$ in the CO Hamiltonian [24,25]:

$$\mathcal{H}_{CDW} = -\sum_{\mathbf{k},\alpha} V_{CDW}(\mathbf{k}) \left( c^\dagger_{\mathbf{k},\alpha} c_{\mathbf{k}+\mathbf{Q}_1,\alpha} + c^\dagger_{\mathbf{k}+\mathbf{Q}_1,\alpha} c_{\mathbf{k},\alpha} \right). \quad (2)$$

Similarly, for disorder-pinned spin density waves (SDW) with a coupling strength $g$ between disorder and SDW [16], we have a CO wave-vector $\mathbf{Q}_2 = \mathbf{Q}_1/2$ [16]:

$$\mathcal{H}_{SDW}^{pinned} = -g^2 \sum_{\mathbf{k},\alpha} V_{SDW}(\mathbf{k}) \left( c^\dagger_{\mathbf{k},\alpha} c_{\mathbf{k}+\mathbf{Q}_2,\alpha} + c^\dagger_{\mathbf{k}+\mathbf{Q}_2,\alpha} c_{\mathbf{k},\alpha} \right).$$
(3)

On the other hand, in the case of commensurate SDW as the relevant CO, the SDW wave-vector becomes $\mathbf{Q}_3 = (\pi,\pi)$, and the corresponding CO Hamiltonian is given by [27]:

$$\mathcal{H}_{SDW} = -\sum_{\mathbf{k},\alpha,\beta} V_{SDW}(\mathbf{k}) \left( c^\dagger_{\mathbf{k}+\mathbf{Q}_3,\alpha} \sigma^3_{\alpha\beta} c_{\mathbf{k},\beta} + c^\dagger_{\mathbf{k},\alpha} \sigma^3_{\alpha\beta} c_{\mathbf{k}+\mathbf{Q}_3,\beta} \right)$$
(4)

where $\sigma^3_{\alpha\beta}$ denotes the matrix element $\alpha\beta$ of the Paul matrix $\sigma^3$.

By incorporating realistic bandstructures and Fermi energies for different families of cuprates with given doping and by specifying the SC pairing symmetry and the form factor for the CO, we can diagonalize $\mathcal{H}_{MF}$ to obtain the bare Green's function $G_0(\mathbf{k},\omega)$ for momentum $\mathbf{k}$ and energy $\omega$. We may further include quantum phase fluctuations between the CO and SC and then solve the Dyson's equation self-consistently for the full Green's function $G(\mathbf{k},\omega)$ [24,25], which gives the quasiparticle spectral density function $A(\mathbf{k},\omega) = -\text{Im}\,[G(\mathbf{k},\omega)]/\pi$ for comparison with ARPES and the quasiparticle density of states $\mathcal{N}(\omega) = \sum_\mathbf{k} A(\mathbf{k},\omega)$ for comparison with STM [24-26].

Based on the aforementioned approach and the assumptions of $d_{x^2-y^2}$-wave pairing and a Gaussian momentum distribution for the CO, the quasiparticle spectra can be fully determined by the following parameters: $\Delta_{SC}$, $V_{CO}$, $\mathbf{Q}$, $\delta\mathbf{Q}$, $\Gamma_\mathbf{k}$ (the quasiparticle linewidth), and $\eta$ (the magnitude of quantum phase fluctuations), which is proportional to the mean-value of the velocity-velocity correlation function [24,28,29]. Our approach leads to the following findings for $0 \leq T \ll T_c$ [24,25]: First, for $V_{CO} > \Delta_{SC}$ and $T = 0$, we obtain two sets of spectral peak features at $\omega = \pm\Delta_{SC}$ and $\omega = \pm\Delta_{eff}$, where $\Delta_{eff} \equiv [(\Delta_{SC})^2 + (V_{CO})^2]^{1/2}$ [24]. Second, the features at $\omega = \pm\Delta_{SC}$ diminish in spectral weight and shift to smaller values with increasing temperature, and eventually vanish at $T_c$ [24]. In contrast, the features at $\omega = \pm\Delta_{eff}$ evolve with temperature into rounded "humps" at $\omega \sim \pm V_{CO}$ for $T \sim T_c$ [24], consistent with the PG phenomena. Third, for $V_{CO} < \Delta_{SC}$, $T^* < T_c$ and $T \ll T_c$, only one set of peaks can be resolved at $\omega = \pm\Delta_{eff}$ and no PG is observed above $T_c$, consistent with the findings in electron-type cuprates [24]. Fourth, in addition to CDW and disorder-pinned SDW, we have explored CO's with $\mathbf{Q}$ other than those along the Cu-O bonding directions [18,27]. We find that quasiparticle spectra resulting from the latter are not compatible with experimental data of the *hole*-type cuprates [24,25]. Fifth, applying our analysis to Bi$_2$Sr$_2$CaCu$_2$O$_x$ (Bi-2212) and YBa$_2$Cu$_3$O$_{7-\delta}$ (Y-123) systems of varying doping levels reveals that the doping dependence of $\Delta_{SC}$ is non-monotonic as that of $T_c$, whereas $V_{CO}$ increases monotonically with decreasing doping [25]. Finally, the quasiparticle lifetime exhibits ``dichotomy" in the hole-type cuprates [24], consistent with experiments [3,7,30].

To examine the applicability of the CO scenario to the Fermi arcs observed in ARPES at $T_c < T < T^*$, we assume that the occurrence of CO below $T^*$ introduces a correlation length $\xi_{CO}$, similar to the superconducting coherence length $\xi_{SC}$ below $T_c$. A finite $\xi_{CO}$ value at $T < T^*$ may be viewed as the result of CO interacting with disorder, which leads to broadening of the CO wave-vector $\mathbf{Q}$ so that $(\xi_{CO})^{-1} \propto |\delta\mathbf{Q}|$. Therefore, for a second-order transition at $T^*$ we expect $|\delta\mathbf{Q}(T)| = \delta\mathbf{Q}(0)\,[1-(T/T^*)]^\nu$ for $T < T^*$ where $\nu$ is a critical exponent. The decreasing $\delta\mathbf{Q}(T)$ with increasing temperature therefore implies weakened disorder pinning potential on the CO. For hole-type cuprates we further restrict $\mathbf{Q}$ to the $(\pi,0)/(0,\pi)$ directions based on the fouth finding outlined above [25]. Thus, we perform fitting to the ARPES dispersion data in Ref. [8] by considering a $T$-independent $\mathbf{Q}$ and $T$-dependent $|\delta\mathbf{Q}|$, and we obtain a set of best fitting parameters ($\Delta_{SC}$, $V_{CO}$, $\mathbf{Q}$, $\delta\mathbf{Q}$) using the temperature Green's function, where we neglect the quantum fluctuations because $T \gg 0$. The consistency of our analysis can be verified by using the ARPES fitting parameters to compute $\mathcal{N}(\omega)$ and then comparing the results with STM data [31].

In Figs. 1(a) – (c) we illustrate the effective gap $\Delta_{eff}(\mathbf{k})$ vs. $\mathbf{k}$ in the two-dimensional Brillouin zone (BZ) of Bi-2212 with three doping levels, where the symbols correspond to the ARPES dispersion data in Ref. [8] and the solid lines are our theoretical curves with best fitting parameters ($\Delta_{SC}$, $V_{CO}$, $\mathbf{Q}$, $\delta\mathbf{Q}$). The temperature dependent $\Delta_{SC}$ and $V_{CO}$ values for three different doping levels in Figs. 1(a) – (c) are illustrated in Figs. 2(a) – (c), showing slight increase of $V_{CO}$ near $T_c$. It is interesting to note that similar behavior has been found in the self-consistent solutions to the gap equations of coexisting SC and CDW under the assumption of phonon-mediated $s$-wave SC and CDW [32]. In Fig. 2(d) we show a simulation based on the assumption of phonon-mediated coexisting SC/CDW detailed in Ref. [32]. While qualitatively similar behavior to the experimental fitting parameters in Figs. 2(a) – 2(c) can be reproduced by finding self-consistent solutions to the gap equations, the phonon cutoff energy $\omega_D = (64.0\pm0.5)$ meV and the phonon induced attractive electron-electron interaction energies $\lambda_\Delta = (328.9\pm1.1)$ meV and $\lambda_G = (672.2\pm2.9)$ meV required for generating SC and CDW gaps comparable to experimental data are found to be unreasonably large relative to known values for the cuprates [33,34], and the resulting $T_c \sim 145$ K is also too high. This finding therefore suggests that the CO scenario can provide qualitatively consistent temperature dependence for the SC and CDW gaps. On the other hand, the underlying mechanism is unlikely dominated by the electron-phonon coupling because of incompatibility of the phonon-mediated mechanism with the $d_{x^2-y^2}$-wave pairing and the large SC and CDW gaps.

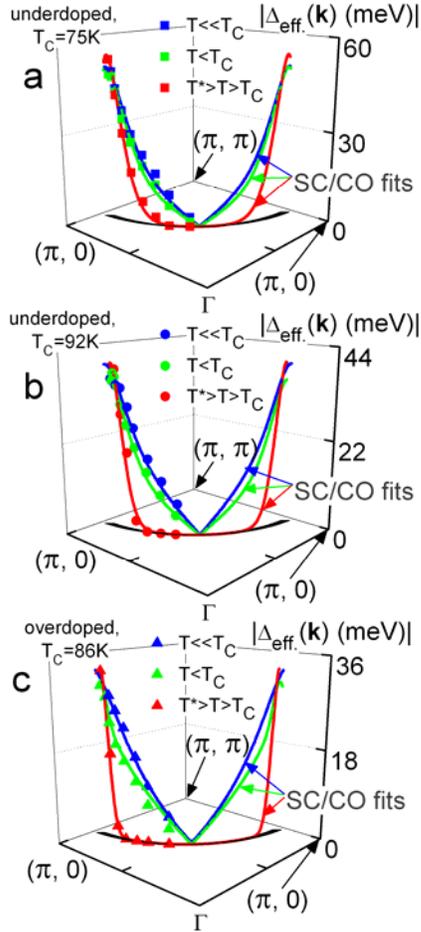

Fig. 1. (Color online) $\Delta_{eff}(\mathbf{k})$ vs. $\mathbf{k}$ of Bi-2212 with three doping levels. The data from Ref. [8] are denoted by the solid symbols, and the fitting curves are given by the solid lines: **(a)** Underdoped sample with $T_c = 75$ K, $T^* = 210$ K, $\delta = 0.11$. The temperature dependent fitting parameters $\Delta_{SC}$ and $V_{CO}$ for $T = 10, 66, 86$ K are given in Fig. 2(a), whereas the CO wave-vector is $|\mathbf{Q}| = 0.16\pi$, and $|\delta\mathbf{Q}| = 0.21\pi$, $0.19\pi$, $0.14\pi$ for $T = 10, 66, 86$ K. **(b)** Slightly underdoped, with $T_c = 92$ K, $T^* = 150$ K, $\delta = 0.15$. The parameters $\Delta_{SC}(T)$ and $V_{CO}(T)$ for $T = 10, 82, 102$ K are given in Fig. 2(b), whereas $|\mathbf{Q}| = 0.2\pi$ and $|\delta\mathbf{Q}| = 0.18\pi, 0.17\pi, 0.1\pi$ for $T = 10, 82, 102$ K. **(c)** Overdoped, with $T_c = 86$ K, $T^* = 100$ K, $\delta = 0.19$. The parameters $\Delta_{SC}(T)$ and $V_{CO}(T)$ for $T = 18, 73, 93$ K are given in Fig. 2(c), whereas $|\mathbf{Q}| = 0.18\pi$ and $|\delta\mathbf{Q}| = 0.22\pi, 0.08\pi, 0.06\pi$ for $T = 18, 73, 93$ K. Both $\mathcal{H}_{CDW}$ and $\mathcal{H}_{SDW}^{pinned}$ yield comparable results. The parameters here are for CDW and for fittings to the anti-bonding band [25], and $\mathbf{Q} \parallel \delta\mathbf{Q} \parallel (\pi,0)/(0,\pi)$.

In addition to the temperature dependent $\Delta_{SC}$ and $V_{CO}$, the $\delta\mathbf{Q}(T)$ values derived from the ARPES data are shown in Fig. 3(a) as a function of $(T/T^*)$, and the doping dependent $T^*$ values for the three samples considered in Ref. [8] are obtained from Ref. [35]. We find a power-law temperature dependence $\delta\mathbf{Q}(T) = \delta\mathbf{Q}(0) [1-(T/T^*)]^\nu$ with $\nu \sim 0.53$, consistent with the mean-field behavior. Next, we use the best fitting parameters to compute the Fermi arc length by finding the momentum interval over which $\Delta_{eff}(\mathbf{k}) = 0$. In Fig. 3(b) we compare the resulting $T$-dependent arc lengths (solid symbols) with a collection of empirical data taken on different hole-type cuprates (crosses) in Ref. [6]. The agreement of our result with the general $(T/T^*)$-dependence of other cuprates implies that our assumption of the Fermi arc being related to the CO correlation length (and therefore the parameter $\delta\mathbf{Q}$) is compatible with experimental findings.

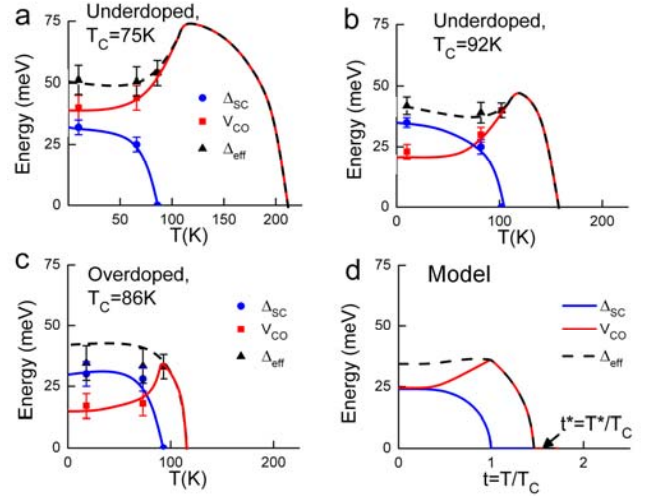

Fig. 2. (Color online) Temperature evolution of $\Delta_{SC}$ and $V_{CO}$ derived from the fitting parameters in Fig. 1 and comparison with the self-consistent solutions in Ref. [32]: **(a)** Fitting parameters for the underdoped sample in Fig. 1(a): $\Delta_{SC} = (32\pm3), (25\pm3), 0$ meV and $V_{CO} = (40\pm5), (44\pm5), (54\pm5)$ meV for $T = 10, 66, 86$ K. **(b)** Fitting parameters for the slightly underdoped sample in Fig. 1(b): $\Delta_{SC} = (35\pm2), (25\pm3), 0$ meV and $V_{CO} = (23\pm5), (30\pm5), (40\pm5)$ meV for $T = 10, 82, 102$ K for $T = 10, 82, 102$ K. **(c)** Fitting parameters for the overdoped sample in Fig. 1(c): $\Delta_{SC} = (30\pm3), (28\pm3), 0$ meV and $V_{CO} = (17\pm5), (18\pm5), (33\pm5)$ meV for $T = 18, 73, 93$ K. **(d)** Simulation of the self-consistent solutions to the temperature dependent gap equations of $\Delta_{SC}(T)$, $V_{CO}(T)$ and $\Delta_{eff}(T)$, following Ref. [32] by assuming phonon mediated CDW and $s$-wave SC. Using a CDW wave-vector $Q = (0.2\pi,0)$, we find that the phonon-related parameters required to yield gap values comparable to experimental observation are unreasonably large relative to known values of the cuprates [33,34]: phonon cutoff energy $\omega_D = (64.0\pm0.5)$ meV, phonon-induced electron-electron attractive energies associated with SC $\lambda_\Delta = (328.9\pm1.1)$ meV and with CDW $\lambda_G = (672.2\pm2.9)$ meV.

Next, we employ the best ARPES fitting parameters ($\Delta_{SC}$, $V_{CO}$, $\mathbf{Q}$, $\delta\mathbf{Q}$) to compute the quasiparticle density of states $\mathcal{N}(\omega)$, and the resulting spectra for three Bi-2212 samples of different doping levels are shown in Fig. 4(a). For comparison, we show in Fig. 4(b) the spatially averaged STM data for Bi-2212 of comparable doping levels [31]. We find overall good agreement between these two sets of spectra. Furthermore, the doping-dependent gap values $\Delta_{SC}$ and $V_{CO}$ derived from ARPES fitting are also consistent with those derived from fitting STM data [25], as shown in Fig. 4(c). We further note that in slightly overdoped Bi-2212, the condition $T^* > T_c$ still holds empirically so that Fermi arcs are observed even though the fitting parameters reveal that $V_{CO} < \Delta_{SC}$. We therefore conclude that the occurrence of Fermi arcs and PG are primarily due to the condition $T^* > T_c$ rather than $V_{CO} > \Delta_{SC}$, provided that $T^*$ is associated with a CO phase transition. We also find that the ratio of $(V_{CO}/k_B T^*) = 2.0 \pm 0.2$ is nearly independent of doping, whereas $(\Delta_{SC}/k_B T^*)$ decreases with increasing $\delta$, from ~ 4.9 for $\delta = 0.11$ to ~ 4.0 for $\delta = 0.19$.

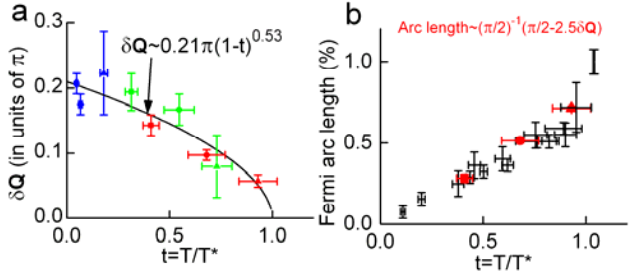

Fig. 3. (Color online) **(a)** $|\delta\mathbf{Q}|$-vs.-$(T/T^*)$ data for Bi-2212 of three different doping levels $\delta = 0.11$ (■), 0.15 (●) and 0.19 (▲) are best fitting results to the ARPES dispersion data in Ref. [8], where the colors are correlated with the temperatures in Fig. 1, and the corresponding $T^*(\delta)$ values are respectively 210 K, 150 K and 100 K according to Ref. [35]. We find the power-law dependence $|\delta\mathbf{Q}(T)| = \delta\mathbf{Q}(0)[1-(T/T^*)]^\nu$, with $\nu \sim 0.53$. **(b)** Fermi arc length vs. $(T/T^*)$, computed from using the $\delta\mathbf{Q}$ values in (a), are denoted by the solid symbols. These values are in agreement with the experimental data (crosses) given in Ref. [6] for other hole-type cuprates.

For completeness, we examine whether the CO scenario with $T^* < T_c$ can account for the absence of Fermi arcs in the electron-type cuprates [24,25]. By assuming $d_{x^2-y^2}$-wave pairing and commensurate SDW with a wave-vector $\mathbf{Q}_3 = (\pi,\pi)$ [27] as the CO for the electron-type cuprate $Pr_{0.89}LaCe_{0.11}CuO_4$ (PLCCO) [36], we employ Eq. (4) to compute the corresponding $\Delta_{eff}(\mathbf{k})$ in the first quadrant of the BZ with $V_{CO} = 4.2$ meV and $\Delta_{SC} = 5.5$ meV. As shown in Fig. 5(a) for $T = 0$ and in Fig. 5(b) for $T = 0.9 T_c$ respectively, we obtain the "Fermi patches" at $T \ll T_c$, and these features evolve into a single gapless point near $T_c$ because SDW has vanished at $T^* < T_c$, which are in good agreement with ARPES data [36]. We further illustrate in Fig. 5(c) the momentum-dependent ARPES leading edge data (×2) from Ref. [36] and the corresponding theoretical fitting curves by assuming either $\mathbf{Q} = (\pi,\pi)$ as in the case of commensurate SDW (dark line) or $\mathbf{Q} \parallel (\pi,0)/(0,\pi)$ as in the case of CDW or disorder-pinned incommensurate SDW (light line). Clearly only the commensurate SDW is consistent with ARPES. The finding that commensurate SDW with a gap $V_{SDW} < \Delta_{SC}$ is a relevant CO in PLCCO has also been corroborated by neutron scattering data on one-layer electron-type cuprates [37].

Finally, we discuss issues associated with representative one-gap models that attribute the formation of Fermi arcs solely to quasiparticle lifetime broadening [11,12]. We note the following difficulties. First, these models attribute the Fermi arcs to an isotropic lifetime broadening above $T_c$, although empirically quasiparticles exhibit dichotomy in their lifetimes at all temperatures [3,7,30]. Second, the one-gap scenario asserts that the occurrence of a finite arc length is due to pair-breaking scattering appearing at $T \geq T_c$. However, thermally induced phase fluctuations that lead to pair-breaking scattering are known to exist even below $T_c$. Moreover, substantial disorder in Bi-2212 may give rise to both quasiparticle and pair-breaking scatterings even for $T \ll T_c$. These effects would have led to occurrence of Fermi arcs below $T_c$, in contradiction to experiments. Third, the assumption of a single $d_{x^2-y^2}$-wave pairing potential cannot be reconciled with our vortex-state quasiparticle spectroscopic studies that revealed PG-like features at $\omega \sim \pm V_{CO}$ inside the vortex core of both electron- and hole-type cuprates at $T \ll T_c$ [25,38] as well as additional field-induced non-dispersive wave-vectors inside vortices [38]; a pure SC phase with $d_{x^2-y^2}$-wave pairing would have led to enhanced spectral weight rather than gapped features inside the vortex core [39]. Moreover, the presence of non-dispersive wave-vectors inside vortices cannot be reconciled with any known theory of pure SC. Finally, the one-gap scenario cannot account for the absence of either Fermi arcs or PG in electron-type cuprates.

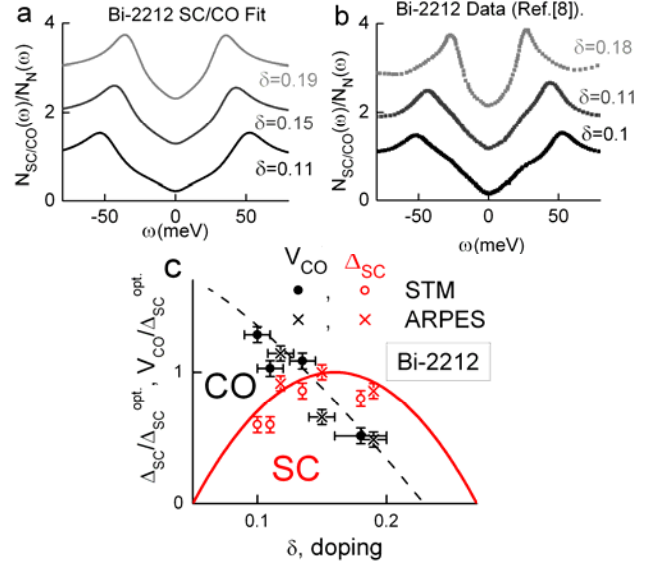

Fig. 4. **(a)** Simulations of the quasiparticle density of states $\mathcal{N}(\omega)$ at $T = 0$ in Bi-2212 of three different doping levels. The input parameters for the simulations are derived from the ARPES fitting in Fig. 1. **(b)** Spatially averaged STM data of three Bi-2212 samples [31], with doping levels comparable to those given in Ref. [8]. **(c)** Comparison of the consistency among the parameters $V_{CO}$ and $\Delta_{SC}$ derived from the ARPES fitting [8] and the STM fitting [25,26]. The solid line represents $T_c(\delta)$ normalized to the optimal doping value, following Ref. [25].

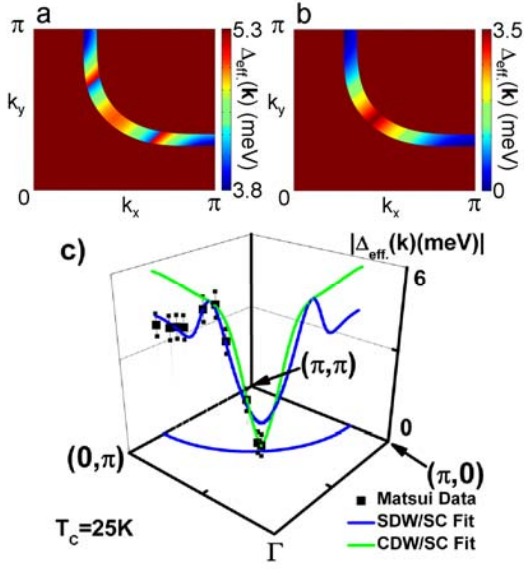

Fig. 5. (Color online) Simulations of $\Delta_{eff}(\mathbf{k})$ in the first quadrant of the BZ of PLCCO at (a) $T = 0$ and (b) $T = 0.9\, T_c$. (c) Momentum dependent ARPES leading edge data ($\times 2$) from Ref. [36] are shown as a function of $\phi \equiv \tan^{-1}(k_y/k_x)$, together with theoretical fitting for two types of CO's, CDW and SDW. Here we have assumed isotropic Gaussian form factors for both $V_{CDW}(\mathbf{k})$ and $V_{SDW}(\mathbf{k})$ in Eqs. (2) and (4). The navy (dark) line corresponds to $\mathbf{Q} = (\pi,\pi)$ for SDW, and the green (light) line corresponds to $\mathbf{Q} \parallel (\pi,0)/(0,\pi)$ for CDW. Clearly the fitting curve with $\mathbf{Q} = (\pi,\pi)$ for SDW agrees much better with ARPES data. Moreover, the presence of commensurate SDW is also consistent with the findings of neutron scattering data on one-layer electron-type cuprates [37].

In summary, we have investigated a possible scenario of CO-induced Fermi arc and PG phenomena in cuprate superconductors above $T_c$. We find that by assuming a CO wave-vector parallel to the antinodal directions, we can account for the *presence* of Fermi arcs and PG phenomena in *hole-type* cuprates as a function of momentum, doping level and temperature if the CO vanishes at a temperature $T^*$ above the SC transition $T_c$. Moreover, we can account for the *absence* of Fermi arcs in *electron-type* cuprates by considering the commensurate SDW with a wave-vector at $(\pi,\pi)$ as the relevant CO that coexists with SC in the ground state and vanishes at a temperature $T^* < T_c$. We have also examined the temperature evolution of the SC and CO gaps in hole-type cuprates by assuming phonon-mediated coexistence of SC and CDW. While the temperature evolution of the gaps agrees *qualitatively* with the solutions to the self-consistent gap equations, we find that the mechanism that leads to coexisting CDW and SC in the hole-type cuprates is unlikely dominated by electron-phonon coupling because of unrealistically large coupling and phonon cutoff energies required to yield results agreeing with empirical observation. Further investigations appear necessary to identify the bosonic mode(s) responsible for mediating the coexistence of CO's and SC in the cuprates and to account for the different CO's among electron- and hole-type cuprates.

**Acknowledgements**

This work was supported by NSF Grant DMR-0405088.